# Mathematical Modeling to Study the Dynamics Of A Diatomic Molecule $N_2$ in Water

Nitin Sharma and Madhvi Shakya

**Abstract**— In the present work an attempt has been made to study the dynamics of a diatomic molecule $N_2$ water. The proposed model consists of Langevin stochastic differential equation whose solution is obtained through Euler's method. The proposed work has been concluded by studying the behavior of statistical parameters like variance in position, variance in velocity and covariance between position and velocity. This model incorporates the important parameters like acceleration, intermolecular force, frictional force and random force.

## 1 Introduction

Molecular dynamics (MD) simulation has been widely used to study the structural and dynamic properties of molecular systems. But, there exist at least two limitations': the approximation in the potential energy functions and the lengths of the simulations. To prolong the simulation time, stochastic dynamics (SD) based on the Langevin equation is a recommended approach in which only the relevant portion of the molecule is considered explicitly and the remainder of system such as solvent serves to provide an effective potential, a friction drag and a random force [1].

The focus of this study is on mathematical modeling of a diatomic molecule $N_2$ for molecular dynamics study in water by using an appropriate and suitable potential energy function for Euler's method. To study the dynamics or motion of the molecule in fluid we have determined and studied the behavior of important statistical parameters like variance in position, variance in velocity and covariance between position and velocity with respect to time respectively.

We used Euler's method for the simulation of the model by replacing the continuous – time with discrete time step, finding values at times $t_0, t_1, \ldots\ldots\ldots\ldots$ usually $t_{n+1} - t_n$ is a fixed number $\Delta t$.

- *Nitin Sharma is with the Maulana Azad National InstituteTechnology, Research Scholar,Department of Mathematics,Bhopal,INDIA*
- *Madhvi Shakya is with the Maulana Azad National InstituteTechnology, ,Department of Mathematics,Bhopal,INDIA*

## 2 Mathematical Model

In the molecular dynamics study, the atoms in a molecule are treated as particles and move according to the Newton's equation of motion by relating intermolecular force and intramolecular force to acceleration [4]. When we incorporate solvent effect then a particle or molecule moving in a fluid medium experiences two additional forces, frictional force due to the viscosity of the medium and random force or fluctuating force due to the collision of the molecule in the surrounding medium then the behavior or dynamics of such molecule can be modeled as [5,10]……..

$$m\ddot{x} = F(x) - \eta\dot{x} + \varepsilon\xi(t) \qquad \ldots (1)$$

Where

$\langle \xi(t)\xi(t') \rangle = \delta(t - t')$, ( $\xi(t) = \frac{dW_t}{dt}$ [6] white noise or random force), $\eta = 6\pi\gamma R$ frictional coefficient or damping force, $F(x) =$ Intermolecular force related to potential function $V(x)$ via $F(x) = \dot{V}(x)$ and $\varepsilon^2 = 2\eta mKT$

Here $\ddot{x}$ and $\dot{x}$ are the second and first derivative of $x$ with respect to time respectively. The term on the left hand side is the product of mass of the molecule $m$ and acceleration $\ddot{x}$. The first term on the right hand side is the inter molecular force, $F(x)$,due to the interaction between the atoms of a molecule. This is the same force used in Newton's equation of motion [7]. And given as,

$$F(x) = -\frac{dV(x)}{dx}$$

Where, $V(x)$ = Intermolecular Potential Energy function.

There are number of potential energy function $V(x)$ for a diatomic molecule, like Harmonic potential, Fourth power potential, Cubic bond stretching potential, Morse potential etc. Some of them in mathematical form can be written as.

**Morse Potential Energy Function**

$$V_{Morse} = D[1 - \exp(-\beta(x-b))]^2$$

Here $D$ is the depth of the well in kj/mol., $\beta$ defines the steepness of the well in ($nm^{-1}$), $b$ is the ideal bond length between two atoms and $x$ is the varying displacement between two atoms[12].

**Harmonic Potential Energy Function**

$$V_{Harmonic} = \frac{1}{2} k_s (x-b)^2$$

Here $k_s$ is Harmonic force constant, $b$ is the ideal bond length between two atoms and $x$ is the varying displacement between two atoms [11].

**Cubic Bond Stretching Potential Energy Function**

$$V_{Cubic} = k_s (x-b)^2 + k_s k_c (x-b)^3$$

Here $k_s$ is Harmonic force constant, $k_c$ Cubic force constant, $b$ is the ideal bond length between two atoms and $x$ is the varying displacement between two atoms [12].

Due to computational and suitability reason of finite difference Euler's method we selected Harmonic potential energy function for simulation of the model (1).Then $F(x)$ can be written as..

$$F(x) = -k_s (x-b)$$

The second term is a frictional force which describes the drag on the particle due to the solvent. This frictional force is proportional to the speed of the particle with the constant of proportionality being the friction coefficient.

$$F_{friction} = -\eta \dot{x}$$

Where $\dot{x}$ is the velocity of the particle and $\eta$ is the friction coefficient [9]. The third term, $\xi(t)$ ,is the random or stochastic force due to thermal fluctuations of the solvent with the amplitude $\varepsilon$ related to the temperature $T$ and the frictional force $\eta$ by the fluctuation dissipation theorem $\varepsilon^2 = 2\eta mKT$ [2].The solvent is not explicitly represented but its effects on the explicit atoms come from the frictional and random forces. When the frictional and random forces are zero, the equation (1) reduces to Newton's equation of motion [9].

Let

$$\frac{dx}{dt} = \frac{dX_t}{dt} = V_t$$

Then

$$m\frac{dV_t}{dt} = F(X_t) - \eta V_t + \varepsilon \frac{dW_t}{dt}$$

Then (1) can be written in terms of $X_t$ and $V_t$ as follows

$$dX_t = V_t dt, \qquad \text{.. (2.1)}$$

$$mdV_t = F(X_t) - \eta V_t + \varepsilon dW_t,$$

In matrix notation above mention equations can be written as...

$$d\begin{pmatrix} X_t \\ V_t \end{pmatrix} = A\begin{pmatrix} X_t \\ V_t \end{pmatrix} dt + \varepsilon \begin{pmatrix} 0 \\ 1/m \end{pmatrix} dW_t + bk_s \begin{pmatrix} 0 \\ 1/m \end{pmatrix} dt,$$

……. (2.2)

Where

$$A = \begin{pmatrix} 0 & 1 \\ -k_s/m & -\eta/m \end{pmatrix}.$$

……. (2.3)

For numerical updates $(2.1)$ can be written as…

$$\begin{pmatrix} X_{n+1} \\ V_{n+1} \end{pmatrix} = C\begin{pmatrix} X_n \\ V_n \end{pmatrix} + \varepsilon B\Delta W_n + Bk_s b\Delta t, \quad \ldots\ldots (2.4)$$

Where

$$C = \begin{pmatrix} c_{11} & c_{12} \\ c_{21} & c_{22} \end{pmatrix}, \quad B = \begin{pmatrix} b_1 \\ b_2 \end{pmatrix}$$

and $\Delta W_n$ is sampled from a Gaussian distribution with mean Zero and Variance $\Delta t$, independently of $\Delta W_n$ for $n \neq m$.

Since $\Delta W_n$ is sampled from Gaussian distribution with mean Zero and Variance $\Delta t$ .So let the covariance matrix of $(2.4)$ be [2].

$$\Sigma = \begin{pmatrix} \sigma_x^2 & \mu \\ \mu & \sigma_v^2 \end{pmatrix},$$

Where $\sigma_x^2, \sigma_v^2$ and $\mu$ are variance in position, variance in velocity and covariance between position and velocity respectively. If $\begin{pmatrix} X_n \\ V_n \end{pmatrix}$ is Gaussian with mean Zero and Covariance matrix $\Sigma$, then the variance of constant term $bk_s B\Delta t$ will be zero and $C\begin{pmatrix} X_n \\ V_n \end{pmatrix}$ is Gaussian with mean Zero and covariance matrix an $C\Sigma C^T$. Thus the covariance matrix that results for $(2.4)$ will be [2].

$$\Sigma = C\Sigma C^T + \varepsilon^2 BB^T \Delta t,$$

Or

$$C\Sigma C^T = \Sigma - \varepsilon^2 BB^T \Delta t, \quad \ldots\ldots\ldots (2.5)$$

We can rewrite $(2.5)$ in a suitable matrix form as follow

$$\begin{pmatrix} c_{11}^2 - 1 & 2c_{11}c_{12} & c_{12}^2 \\ c_{11}c_{12} & c_{11}c_{22} + c_{12}c_{21} - 1 & c_{12}c_{22} \\ c_{21}^2 & 2c_{21}c_{22} & c_{22}^2 - 1 \end{pmatrix} \begin{pmatrix} \sigma_x^2 \\ \mu \\ \sigma_v^2 \end{pmatrix} = -\varepsilon^2 \Delta t \begin{pmatrix} b_1^2 \\ b_1 b_2 \\ b_2^2 \end{pmatrix}.$$

### 3 The Euler's method

By the Euler's method, the position and velocity can be updated as follows [2].

$$X_{n+1} = X_n + V_n \Delta t,$$

$$\ldots\ldots (3.1)$$

$$mV_{n+1} = mV_n - \eta V_n \Delta t + f(X_n)\Delta t + \varepsilon \Delta W_n.$$

This gives the oscillatory motion for position and velocity respectively for the diatomic molecule $N_2$ as

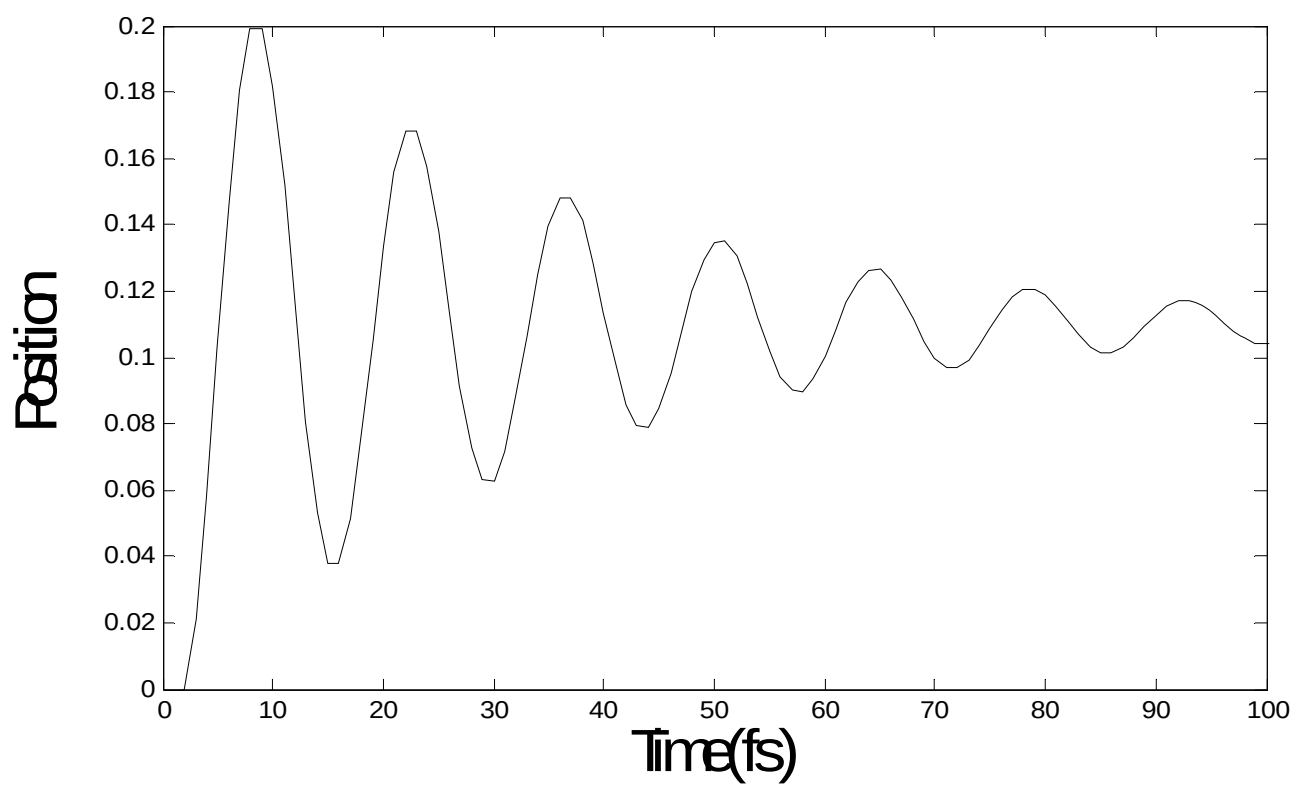

Figure 1

And

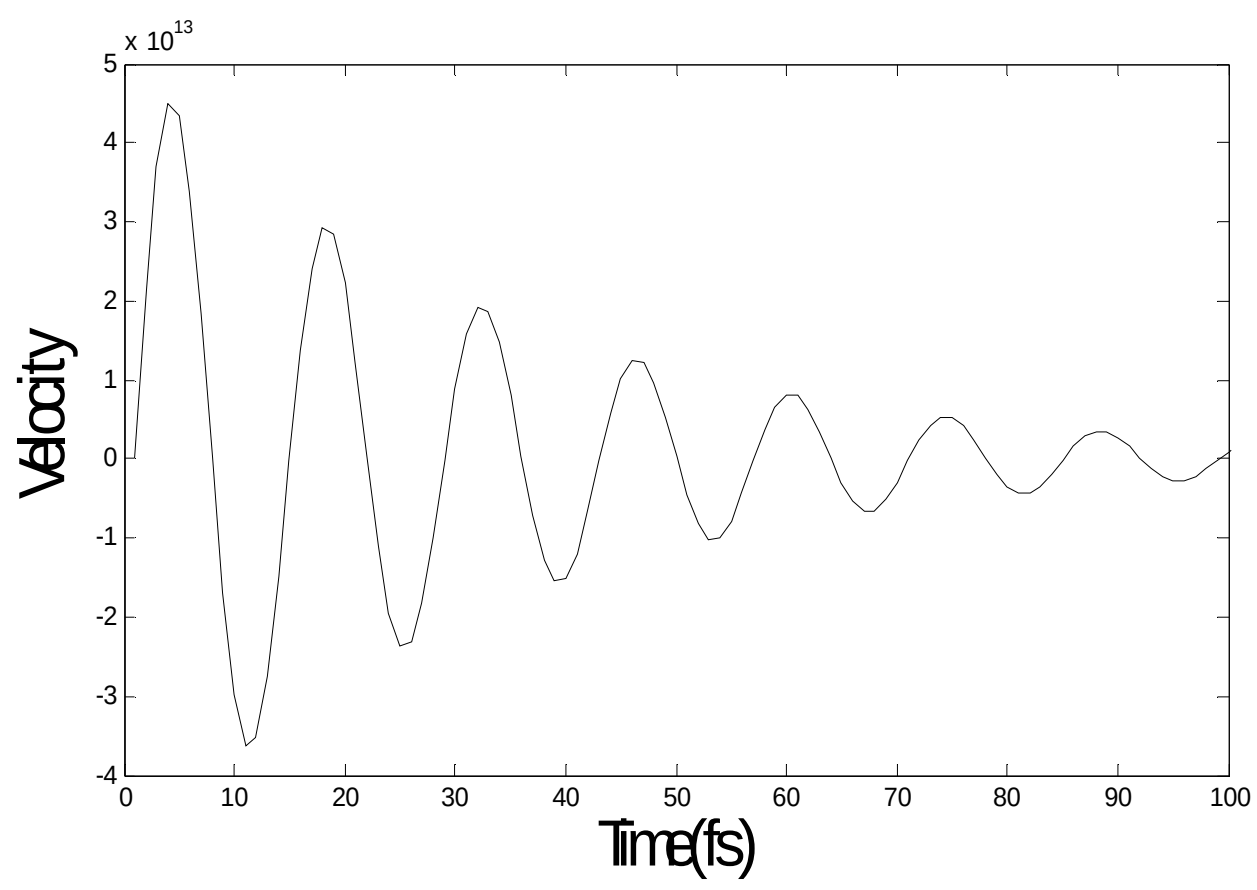

Figure 2

Figure 1 and Figure 2 shows the motion of diatomic molecule $N_2$ in terms of position and velocity up to 100femtosecond for standard wiener process (i.e. for wiener process of mean 0 and variance 1). It is clear from figures (Figure 1 and Figure 2) that oscillations in position and velocity in the motion of diatomic molecule $N_2$ reduces as time increases which indicates the frictional (damping) force effect caused by water due to which oscillations in motion are reducing.

Now with the notation of (2.4) , we have

$$C = I + \Delta t P = \begin{pmatrix} 1 & \Delta t \\ -k_s \Delta t / m & 1 - \eta / m \Delta t \end{pmatrix}, \quad B = \begin{pmatrix} 0 \\ 1/m \end{pmatrix}$$

Then the solution of above equation (2.5) given as

$$\sigma_x^2 = \frac{2\eta KTm(2m - \eta\Delta t + k_s \Delta t^2)}{(\eta - k_s \Delta t)(4mk_s - 2k_s \eta \Delta t + k_s^2 \Delta t^2)}$$

$$\sigma_v^2 = \frac{4\eta KTmk_s}{(\eta - k_s \Delta t)(4mk_s - 2k_s \eta \Delta t + k_s^2 \Delta t^2)},$$

And $$\mu = \frac{-2\eta KTm\Delta t k_s}{(\eta - k_s \Delta t)(4mk_s - 2k_s \eta \Delta t + k_s^2 \Delta t^2)}$$

## 4 Numerical Results And Discussion

In the present study we done modeling and simulation for a diatomic molecule $N_2$ and computed the important parameters, like variance in position, variance in velocity and covariance between position and velocity to study the dynamic aspects of diatomic molecule. We have observed that variance in position (Figure 3) increases steadily upto $1.29 \times 10^{-15}$sec after that suddenly it increases, decreases and then increases unexpectedly from the point $1.30 \times 10^{-15}$sec. to $1.32 \times 10^{-15}$sec. i.e. it starts showing the chaotic behavior from the time steps $1.30 \times 10^{-15}$sec. to $1.32 \times 10^{-15}$sec. and then it become stable. Similarly, we have computed variance in velocity (Figure 4) it increases steadily up to $1.29 \times 10^{-15}$sec. after that suddenly it increases, decreases and then increases unexpectedly from the point $1.30 \times 10^{-15}$sec. to $1.32 \times 10^{-15}$sec. i.e. again it starts showing the chaotic behavior for the same time steps $1.30 \times 10^{-15}$sec.. to $1.32 \times 10^{-15}$sec. and then it become stable. Finally we have computed the covariance (Figure 5) between position and velocity which also increases steadily up to $1.29 \times 10^{-15}$sec. after that again suddenly it increases, decreases and then increases unexpectedly again from the point $1.30 \times 10^{-15}$sec. to $1.32 \times 10^{-15}$sec. i.e. again it starts showing the chaotic behavior from the same time Steps $1.30 \times 10^{-15}$sec. to $1.32 \times 10^{-15}$sec. and then it become stable From these three observations of statistical parameter we have concluded that the motion of the molecule $N_2$ becomes unstable after $1.29 \times 10^{-15}$sec . and then it starts showing the chaotic behavior from the time step $1.30 \times 10^{-15}$sec. to $1.32 \times 10^{-15}$sec.. Also it is observed that Euler's method is unstable, if the condition of stability is not satisfied i.e. $k_s \Delta t^2 < \eta \Delta t$ or $\eta \Delta t < 2 + \frac{1}{2} k_s \Delta t^2$ [2].

We have taken a diatomic molecule $N_2$ in water and for this molecule we have taken the values for the parameters incorporated in above equations as follow [7, 8, 3, and 11]

$m = 1.16265 \times 10^{-23}\,\text{pN.}\dfrac{\text{sec}^2}{\text{nm}}$,

$\eta = 2.9229 \times 10^{-9}\,\text{pN. sec./nm.}$,

$KT = 4.1\text{pN.nm(at room temprature).}$

$k_s = 2240000\ \text{pN./nm.}$

First, we have calculated and plotted variance in position against time step $\Delta t$ up to $4\times10^{-15}$sec.

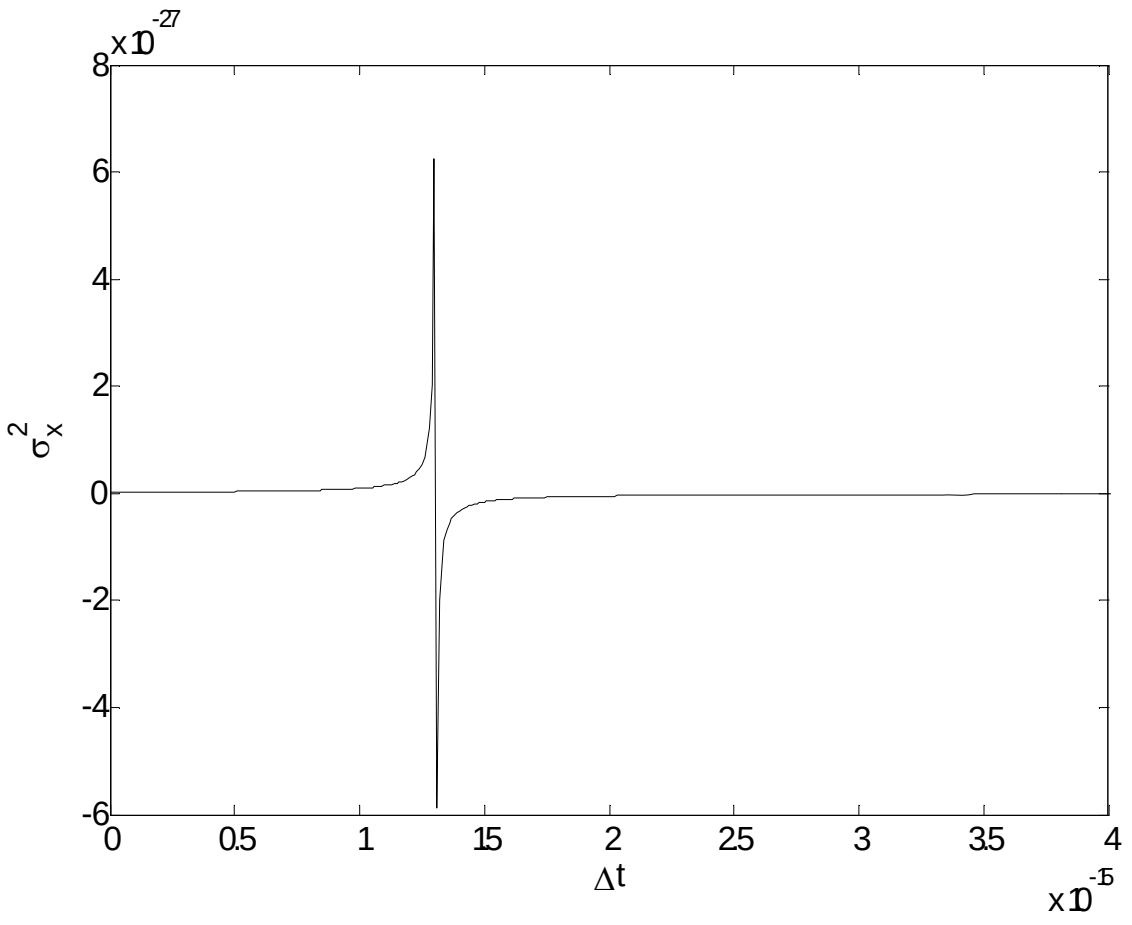

Figure 3

Now, we have calculated variance in velocity against time step $\Delta t$ up to $4\times10^{-15}$sec.

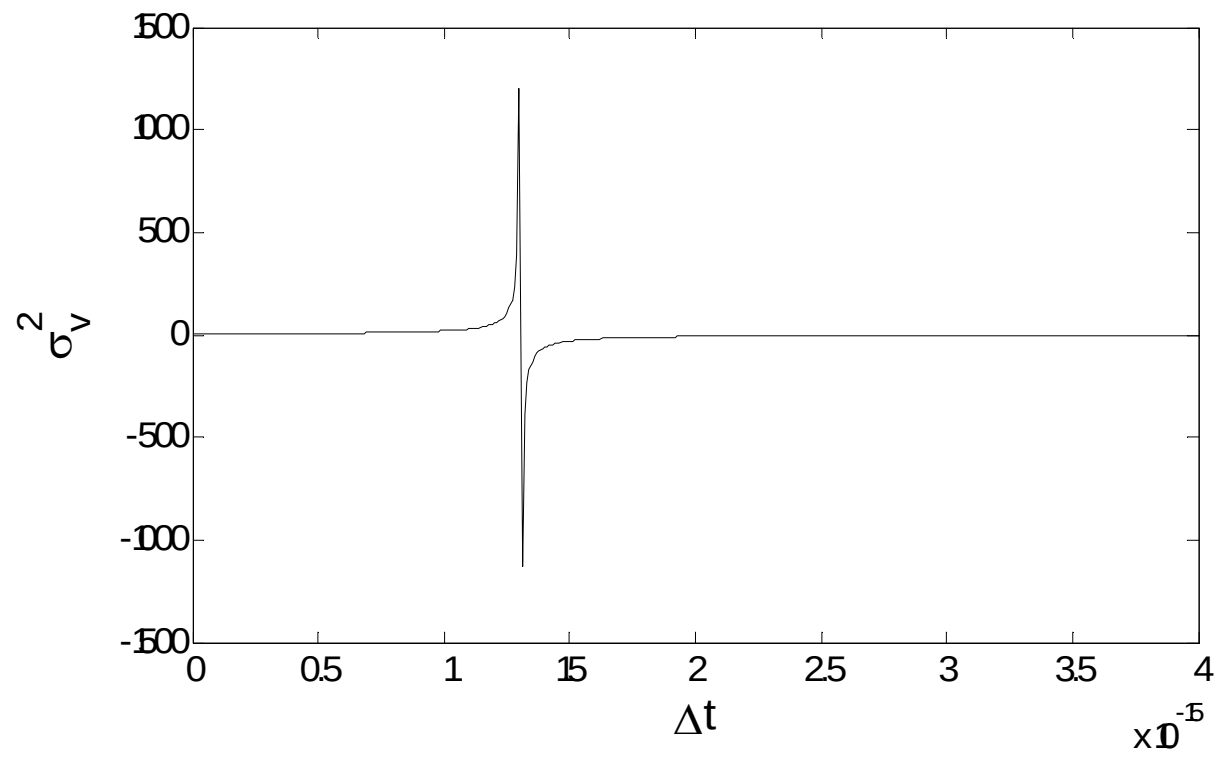

Figure 4

Finally, we have calculated and plotted covariance between position and velocity against time step $\Delta t$ upto $4\times10^{-15}$sec .

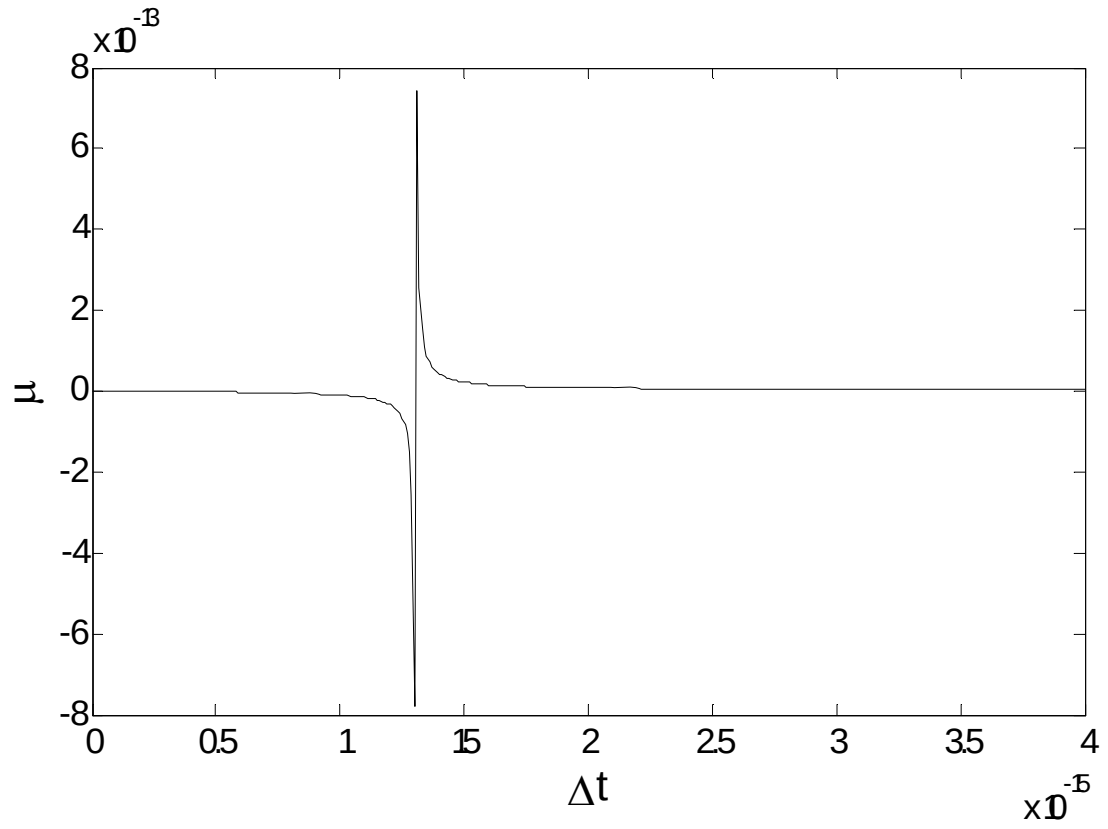

Figure 5

## 5 CONCLUSION

In this paper an attempt has been made for mathematical modeling to study the dynamics of a diatomic molecule $N_2$ by incorporating the solvent effect. We have used Euler's method for simulation and we have obtained the simulated results up to time step $10^{-15}$ sec, it is observed that Euler's method is stable for the condition $k_s\Delta t^2 < \eta\Delta t$ or $\eta\Delta t < 2 + \frac{1}{2}k_s\Delta t^2$ [2].

Also, We have observed by studying the behavior of variance in position, variance in velocity and covariance between velocity and position that motion or dynamics of diatomic molecule $N_2$ remains stable up to time step $1.29\times10^{-15}$ sec . and after that it become unstable and start showing the chaotic behavior.

**Acknowledgements.** We are grateful to Department of Biotechnology, New Delhi and MP Council of Science and Technology, Bhopal for providing infrastructure facility for MANIT, Bhopal for carrying out this research work.

**NitinSharma** did his M.Sc. (Mathematics) From SOMAAS, Jiwaji University Gwalior, India (2006) and at present pursuing his Ph.D. in Applied Mathematics from Maulana Azad National Institute Of Technology,Bhopal,India

**Madhvi Shakya** is Assistant Professor in the Department of Mathematics, Maulana Azad National Institute Of Technology, Bhopal (M.P.), India. She obtained her M.Sc. (Mathematics) from Jiwaji University, Gwalior (M.P.), India and Ph.D. in Mathematics from Rajiv Gandhi Proudyogiki Vishwavidyalaya (M.P. state Technological University), Bhopal,India. Her research interests include computational and Mathematical Biology, Bioinformatics, Biocomputing/Biomodeling.